%% file: HIG-16-017_temp.tex
\pdfoutput=1

\documentclass[11pt,twoside,a4paper,cmspaper,final,collab]{cms-tdr}
\def\svnVersion{485032P}\def\cmsCernNoTag{CERN-EP-2018-204}\def\cmsCernDate{\today}\def\cmsMessage{Published in the Journal of High Energy Physics as \href{http://dx.doi.org/10.1007/JHEP11(2018)161}{\doi{10.1007/JHEP11(2018)161.}}}

\begin{document}\cmsNoteHeader{HIG-16-017}

\hyphenation{had-ron-i-za-tion}
\hyphenation{cal-or-i-me-ter}
\hyphenation{de-vices}
\RCS$HeadURL: svn+ssh://svn.cern.ch/reps/tdr2/papers/HIG-16-017/trunk/HIG-16-017.tex $
\RCS$Id: HIG-16-017.tex 482088 2018-11-22 13:43:28Z anikiten $

\providecommand{\NA}{\ensuremath{\text{---}}\xspace}
\newcommand{\PX}{\ensuremath{\cmsSymbolFace{X}}\xspace} 
\newcommand{\mX}{\ensuremath{m_\PX}\xspace} 
\newcommand{\mmm}{\ensuremath{m_{\PGm\PGm}}\xspace}
\newcommand{\Gmm}{\ensuremath{\Gamma_{\PGm\PGm}}\xspace}
\newcommand{\NS}{\ensuremath{N_\mathrm{S}}\xspace}
\newcommand{\NB}{\ensuremath{N_\mathrm{B}}\xspace}
 \cmsNoteHeader{HIG-16-017}
\title{Search for resonances in the mass spectrum of muon pairs produced in association
       with {\cPqb} quark jets in proton-proton collisions at $\sqrt{s} = 8$ and 13\TeV}

\date{\today}
\abstract{A search for resonances in the mass range 12--70\GeV produced in association with a \PQb quark jet and a second jet, and decaying to a muon pair, is reported. The analysis is based on data from proton-proton collisions at center-of-mass energies of 8 and 13\TeV, collected with the CMS detector at the LHC and corresponding to integrated luminosities of 19.7 and 35.9\fbinv, respectively. The search is carried out in two mutually exclusive event categories. Events in the first category are required to have a \PQb quark jet in the central region ($\abs{\eta}\le 2.4$) and at least one jet in the forward region ($\abs{\eta}>2.4)$. Events in the second category are required to have two jets in the central region, at least one of which is identified as a \PQb quark jet, no jets in the forward region, and low missing transverse momentum. An excess of events above the background near a dimuon mass of 28\GeV is observed in the 8\TeV data, corresponding to local significances of 4.2 and 2.9 standard deviations for the first and second event categories, respectively. A similar analysis conducted with the 13\TeV data results in a mild excess over the background in the first event category corresponding to a local significance of 2.0 standard deviations, while the second category results in a 1.4 standard deviation deficit. The fiducial cross section measurements and 95\% confidence level upper limits on those for a resonance consistent with the 8\TeV excess are provided at both collision energies.}
\hypersetup{
pdfauthor={CMS Collaboration},
pdftitle={Search for narrow resonances in the mass spectrum of muon pairs produced in association with b quark jets},
pdfsubject={CMS},
pdfkeywords={CMS, Higgs physics, Beyond Standard Model}}
\maketitle
\section{Introduction}
\label{sec:introduction}

The discovery of a Higgs boson~\cite{ATLAS-Higgs,CMS-Higgs1,CMS-Higgs2} with a mass near 125\GeV~\cite{ATLAS-CMS-mass,CMS-HZZ} provided new motivation to search for an extended Higgs sector at the CERN LHC. These searches are focused not only on additional Higgs bosons at high mass, but also on possible light states below 125\GeV, which may have eluded earlier detection. Light (pseudo)scalar bosons are predicted in a number of beyond the standard model (SM) theories, e.g., in two Higgs doublet models (2HDM)~\cite{Branco:2011iw} and next-to-minimal supersymmetric SM (NMSSM)~\cite{Ellwanger:2009dp}. Despite an extensive program of searches for such resonances by the CERN LEP experiments~\cite{Schael:2010aw,Schael:2006cr,Abbiendi:2004ww,Abdallah:2007ab,Abbiendi:2002in}, and by the ATLAS~\cite{Aad:2015oqa,Aad:2015bua,Aaboud:2016oyb,Aaboud:2018fvk,Aaboud:2018gmx,Aaboud:2018iil,Aaboud:2018esj}, CMS~\cite{Chatrchyan:2012am,CMS:2015ooa,Khachatryan:2015nba,Khachatryan:2015baw,Khachatryan:2015wka,Khachatryan:2017mnf,Sirunyan:2017tqo,Sirunyan:2017uvf}, and LHCb~\cite{Aaij:2013nba} Collaborations, the present experimental limits on the product of production cross sections and branching fractions do not yet exclude the existence of such particles~\cite{Aggleton:2016tdd}.

As numerous searches for heavy particles at the LHC have thus far produced only null results, searches for low-mass resonances with suppressed couplings to SM particles have received increased interest. Examples include extending dijet resonance searches to low masses~\cite{CMS-ISR1,CMS-ISR2,CMS-ISR3,ATLAS-ISR}, and searches for dark photons and dark \PZ bosons~\cite{Aad:2015sva,Aaij:2017rft,Aaboud:2018fvk}. Such low-mass resonances are predicted in a number of models, including those~\cite{Dalchenko:2017shg, Kamenik:2017tnu} providing possible explanations for the host of recently observed flavor anomalies~\cite{Aaij:2014ora,Aaij:2017vbb} via $\PZpr$ bosons with nonuniversal couplings to quarks and leptons. The cross section of the associated production with bottom quarks of a new light boson (scalar or vector), times the dimuon branching fraction of its decay, can be large in proton-proton (\Pp\Pp) collisions at the LHC, e.g., in 2HDM~\cite{Bernon:2014nxa} or in $\PZpr$~\cite{Dalchenko:2017shg} models. Previous searches in this channel were performed by CMS using $\sqrt{s} = 7$ and 8\TeV data~\cite{CMS:2015ooa,Sirunyan:2017uvf}.

In the course of detailed studies related to a search~\cite{Sirunyan:2017uvf} for a (pseudo)scalar boson produced in association with bottom quarks and decaying into opposite-sign (OS) muon pairs, $\Pp\Pp \to \bbbar \PSA$, $\PSA \to \PGmp\PGmm$, performed by CMS at a center-of-mass energy of $\sqrt{s}=8\TeV$ in 2012, an enhancement in the dimuon spectrum near 28\GeV was observed in events containing a {\cPqb} quark jet ("\PQb jet") in the central pseudorapidity region ($\abs{\eta} \le 2.4$) and another jet in the forward region ($\abs{\eta} > 2.4$). The excess is vanishing in the published analysis~\cite{Sirunyan:2017uvf}, being diluted by much more inclusive selections applied to data. As a cross-check, a complementary sample of events with two OS muons, a central \PQb jet, an additional central jet, no forward jets, and low missing transverse momentum was studied. An excess of events above the SM background was observed also in this independent sample. Extensive studies related to various features of the observed excess and its possible origin did not reveal any significant systematic biases or problems with the background estimation methods or with the analysis technique. A similar analysis has now been performed using data collected in 2016 at $\sqrt{s}=13\TeV$. It results in a mild excess over the background in the first event category corresponding to a local significance of 2.0 standard deviations (s.d.), while the second category results in a 1.4\unit{s.d.} deficit.

This paper describes in detail both the 8 and 13\TeV analyses, corresponding to integrated luminosities of 19.7\ and 35.9\fbinv, respectively, and is organized as follows. The CMS detector is briefly described in Section~\ref{sec:detector}. Data and simulated samples, as well as the event reconstruction, are presented in Section~\ref{sec:reco}. The event selection is described in Section~\ref{sec:event}, followed by a statistical characterization of the observed dimuon mass distributions in Section~\ref{sec:results}. Results are summarized in Section~\ref{sec:summary}.

\section{The CMS detector\label{sec:detector}}

The central feature of the CMS apparatus is a superconducting solenoid of 6\unit{m} internal diameter, providing a magnetic field of 3.8\unit{T}. Within the solenoid volume are a silicon pixel and strip tracker, a lead tungstate crystal electromagnetic calorimeter (ECAL), and a brass and scintillator hadron calorimeter (HCAL), each composed of a barrel and two endcap sections. Forward calorimeters extend the pseudorapidity coverage provided by the barrel and endcap detectors. Muons are detected in gas-ionization chambers embedded in the steel flux-return yoke outside the solenoid.

In the region $\abs{\eta} < 1.74$, the HCAL cells have widths of 0.087 in pseudorapidity and 0.087 in azimuth ($\phi$). In the $\eta$--$\phi$ plane, and for $\abs{\eta} < 1.48$, the HCAL cells map on to $5 \times 5$ arrays of ECAL crystals to form calorimeter towers projecting radially outwards from close to the nominal interaction point. For $\abs{\eta} > 1.74$, the coverage of the towers increases progressively to a maximum of 0.174 in $\Delta \eta$ and $\Delta \phi$. Within each tower, the energy deposits in ECAL and HCAL cells are summed to define the calorimeter tower energies, subsequently used to provide the energies and directions of hadronic jets. 

Muons are measured in the pseudorapidity range $\abs{\eta} < 2.4$, with detection planes made using three technologies: drift tubes, cathode strip chambers, and resistive-plate chambers.

Events of interest are selected using a two-tiered trigger system~\cite{Khachatryan:2016bia}. The first level, composed of custom hardware processors, uses information from the calorimeters and muon detectors to select events at a rate of around 100\unit{kHz} within a time interval of less than 4\mus. The second level, known as the high-level trigger, consists of a farm of processors running a version of the full event reconstruction software optimized for fast processing, and reduces the event rate to around 1\unit{kHz} before data storage.

A more detailed description of the CMS detector, together with a definition of the coordinate system used and the relevant kinematic variables, can be found in Ref.~\cite{Chatrchyan:2008aa}.

\section{Data, simulation, and event reconstruction}
\label{sec:reco}

Online, the events were selected by requiring a single-muon trigger with a \pt threshold of 24\GeV, loose isolation requirements, and a fiducial requirement of $\abs{\eta} < 2.1$ for the muon. The trigger efficiency for the events selected for the analysis (Section~\ref{sec:event}) is 95\% for both center-of-mass energies.

The particle-flow (PF) algorithm~\cite{Sirunyan:2017ulk} aims to reconstruct and identify each individual particle in an event, with an optimized combination of information from the various elements of the CMS detector. The energy of muons is obtained from the curvature of the corresponding track. The energy of photons is directly obtained from the ECAL measurement. The energy of electrons is determined from a combination of the electron momentum at the primary interaction vertex, as determined by the tracker, the energy of the corresponding ECAL cluster, and the energy sum of all bremsstrahlung photons spatially compatible with originating from the electron track. The energy of charged hadrons is determined from a combination of their momentum measured in the tracker and the matching ECAL and HCAL energy deposits, corrected for zero-suppression effects and for the response function of the calorimeters to hadronic showers. Finally, the energy of neutral hadrons is obtained from the corresponding corrected ECAL and HCAL energy. The missing transverse momentum vector in the event \ptvecmiss is defined as a negative vectorial sum of the \pt of all PF candidates in an event; its magnitude is referred to as \ptmiss.

For each event, hadronic jets are clustered from PF candidates using the infrared- and collinear-safe anti-\kt algorithm~\cite{Cacciari:2008gp} with a distance parameter of 0.5 (0.4) for the 8 (13)\TeV analysis, as implemented in the \FASTJET package~\cite{Cacciari:2011ma}. The jet momentum is determined as the vectorial sum of all particle momenta in the jet, and is found from simulation to be within 5 to 10\% of the true momentum over the entire \pt spectrum and detector acceptance. Additional \Pp\Pp\ interactions within the same or nearby bunch crossings (pileup) can contribute additional tracks and calorimetric energy depositions to the jet momentum. To mitigate this effect, tracks identified to be originating from pileup vertices are discarded and an offset correction~\cite{Cacciari:2007fd} is applied to correct for remaining contributions. Jet energy corrections are derived from simulation to bring the measured response of jets to that of particle-level jets on average. In situ measurements of the momentum balance in dijet, multijet, photon+jet, and leptonically decaying $\PZ$+jet events are used to account for any residual differences between the jet energy scales in data and simulation~\cite{Khachatryan:2016kdb}. The jet energy resolution amounts typically to 15\% at 10\GeV, 8\% at 100\GeV, and 4\% at 1\TeV. Additional selection criteria are applied to remove jets potentially dominated by anomalous contributions from various subdetector components or reconstruction failures~\cite{CMS-PAS-JME-10-003}. Jets identified as likely coming from pileup~\cite{CMS-PAS-JME-13-005} are also removed.

Jets originating from \PQb quarks are tagged by using multivariate analysis (MVA) algorithms. The CSVMVA~\cite{Chatrchyan:2012jua,CMS-PAS-BTV-13-001} (cMVAv2~\cite{Sirunyan:2017ezt}) algorithm is used in the 8 (13)\TeV analysis. The MVA algorithms take as inputs the impact parameters of jet constituents and secondary vertices reconstructed within the jet~\cite{Chatrchyan:2014fea}. We use the ``tight" working point of the \PQb tagging algorithms at both collision energies, which corresponds to approximately 50\% \PQb jet tagging efficiency and 0.1\% light-quark or gluon jet mistag rate for the jets within the kinematic range used in the analysis. The misidentification rate for \cPqc\ quark jets is 2\%.

Muons are reconstructed using a simultaneous global fit performed with the hits in the silicon tracker and the muon system. They are required to pass standard identification criteria~\cite{Chatrchyan:2013sba,Sirunyan:2018fpa} based on the minimum number of hits in each detector, quality of the fit, and the consistency with the primary vertex, by requiring the longitudinal and transverse impact parameters to be less than 0.5 and 0.2\unit{cm}, respectively. The efficiency to reconstruct and identify muons is greater than 96\%. Matching muons to tracks measured in the silicon tracker results in a relative transverse momentum (\pt) resolution for muons with $20 <\pt < 100\GeV$ of 1.3--2.0\% in the barrel and better than 6\% in the endcaps. The \pt resolution in the barrel is better than 10\% for muons with \pt up to 1\TeV~\cite{Chatrchyan:2012xi}. Muons must be isolated from other activity in the tracker by requiring the \pt sum of other charged PF candidates within a cone of radius $\Delta R =\sqrt{\smash[b]{(\Delta\eta)^2+(\Delta\phi)^2}} = 0.3$, centered on the muon candidate, to be less than 10\% of the muon candidate \pt. If the two muons with the highest \pt in an event are within the isolation cone of one another, the other muon candidate is removed from the isolation sum for each muon.

The reconstructed vertex with the largest value of summed charged-particle track (physics-object) $\pt^2$ is taken to be the primary $\Pp\Pp$ interaction vertex in the 8 (13)\TeV analysis. The physics objects are the jets, clustered using the jet finding algorithm~\cite{Cacciari:2008gp,Cacciari:2011ma}, with the tracks assigned to the vertex as inputs. Events are required to have at least one primary vertex, with the position along (transverse to) the direction of the beams within 24 (2)\unit{cm} of the geometrical center of the detector.

Simulated event samples are used to study the backgrounds. The following background processes were considered: Drell--Yan (DY), W+jets, \ttbar, single top quark, and diboson (VV) production. The DY background includes the associated production of $\ell^{+}\ell^{-}$ ($\ell=\Pe,\mu,\tau$) pairs with \PQc and {\cPqb} quarks. Monte Carlo (MC) simulation of these processes in the 8\TeV analysis is described in detail in Ref.~\cite{Sirunyan:2017uvf}. Events are generated either at leading order (LO) with the \MADGRAPH v5.1.3.30 generator~\cite{Alwall:2011uj} or at next-to-leading order (NLO) with \POWHEG 1.0~\cite{Alioli:2010xd,Nason:2004rx,Frixione:2007vw}. The CTEQ6~\cite{Pumplin:2002vw} parton distribution functions (PDFs) are used in the matrix element calculations. The parton shower and fragmentation are described by \PYTHIA v6.426~\cite{Sjostrand:2006za} with the Z2$^*$ underlying event tune~\cite{Chatrchyan:2013gfi,Khachatryan:2015pea}. In the 13\TeV analysis, we use \MGvATNLO v2.2.2 or higher~\cite{Alwall:2014hca} and \POWHEG v2.0 with NNPDF3.0 PDFs~\cite{Ball:2014uwa}, followed by \PYTHIA v8.212~\cite{Sjostrand:2014zea} with the CUETP8M1 underlying event tune~\cite{Khachatryan:2015pea}. The cross sections of generated samples are normalized to the highest order theoretical calculations available, NLO or higher.

A detector simulation based on \GEANTfour (v.9.4p03 for 8\TeV and v.10.02.p02 for 13\TeV analysis)~\cite{Agostinelli:2002hh} is applied to all generated samples. The effect of pileup is accounted for by superimposing simulated minimum bias events on the hard scattering process, with a multiplicity distribution that matches the one observed in data. The \PQb tagging and muon reconstruction efficiencies, as well as the jet energy scale and resolution in simulation, are corrected to match the corresponding values measured in data.

\section{Event selection\label{sec:event}}

The candidate event selection follows closely that of Ref.~\cite{Sirunyan:2017uvf}. We require an OS muon pair with both muons passing the $\pt > 25\GeV$ and $\abs{\eta} < 2.1$ requirements. The dimuon invariant mass \mmm is required to exceed 12\GeV in order to remove low-mass resonances and poorly modeled backgrounds. We require at least two jets with $\pt > 30\GeV$ and $\abs{\eta} < 4.7$ in an event, with at least one of them found in the central region $\abs{\eta} \le 2.4$ and being {\cPqb} tagged. We further define two search regions (SRs): one with no other central jets (SR1) and one with a second jet found in the central region, no jets in the forward region ($\abs{\eta} > 2.4$),
$\ptmiss < 40\GeV$, and the azimuthal angle between the direction of the dimuon and dijet systems $\Delta\phi(\mu\mu,{\mathrm{jj}}) > 2.5$ radians (SR2). Table~\ref{table:topologies} summarizes the event selection described above.

\begin{table}[htp]
  \centering
  \topcaption{Event selection in the two search regions. A dash means that the variable is not used for selection.}
  \label{table:topologies}
  \begin{tabular}{ c c c }
    \hline
    Event       & SR1  & SR2  \\
    category      & Additional forward jet & Additional central jet\\
     \hline
    Muons & \multicolumn{2}{c}{OS, $\pt>25\GeV, \abs{\eta}<2.1$} \\
   \mmm  & \multicolumn{2}{c}{$\mmm > 12\GeV$}\\
    \cPqb-tagged jet & \multicolumn{2}{c}{$\pt>30\GeV, \abs{\eta}\le 2.4$} \\
    Additional jet & $\pt>30\GeV, 2.4 < \abs{\eta} < 4.7$ &  $\pt>30\GeV, \abs{\eta}\le2.4$ \\
    Jet veto & No other jets $\pt>30\GeV, \abs{\eta}\le2.4$  & No jets $\pt>30\GeV, 2.4<\abs{\eta}<4.7$  \\
    $\ptmiss$ & \NA  & $<$40\GeV \\
    $\Delta\phi(\mu\mu,{\mathrm{jj}})$ & \NA & $>$2.5\unit{rad} \\
    \hline
  \end{tabular}
  \centering
\end{table}

\begin{figure}[tb!]
    \centering
\vspace{1cm}
   \includegraphics[width=0.475\textwidth]{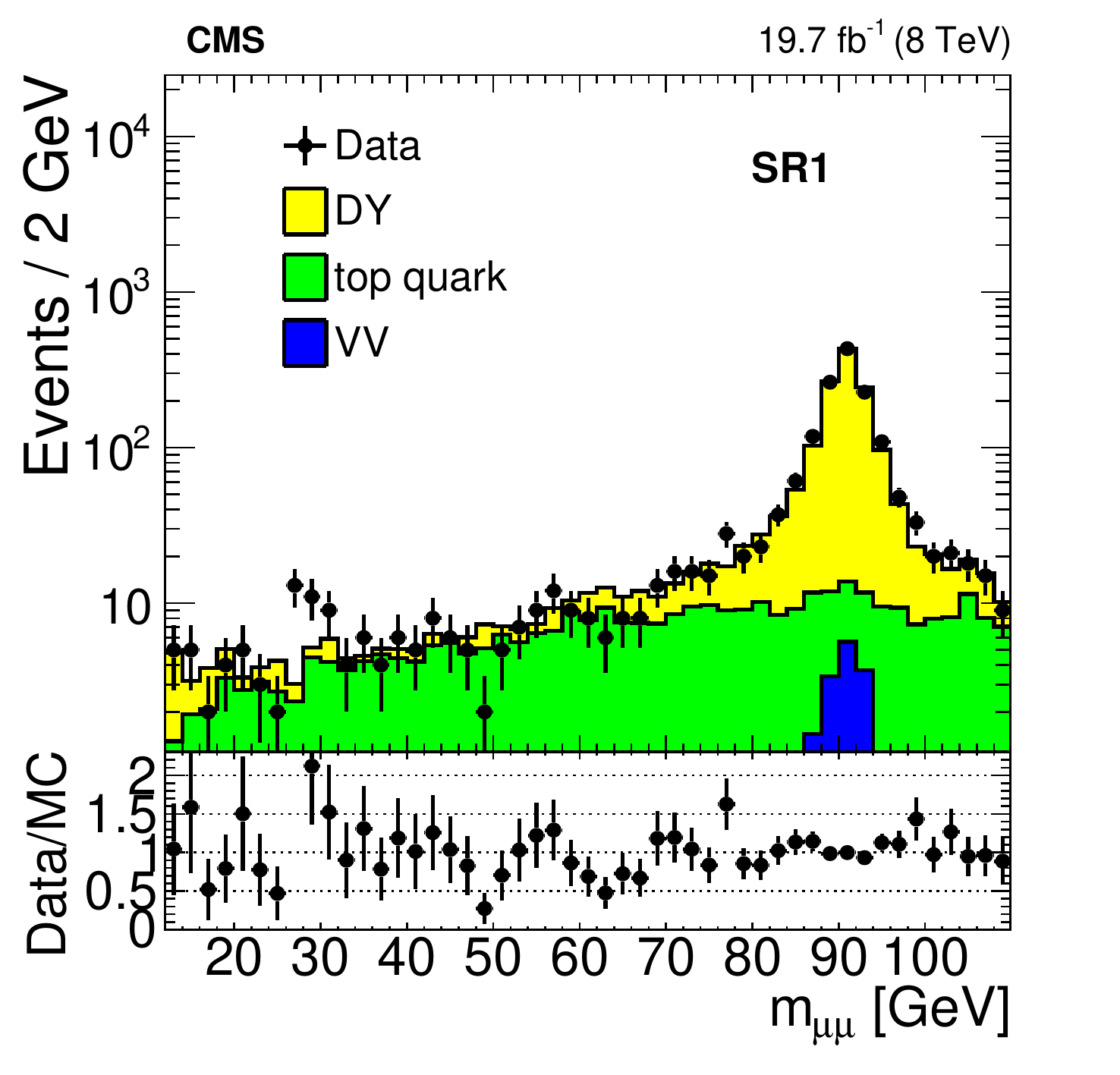}
   \includegraphics[width=0.475\textwidth]{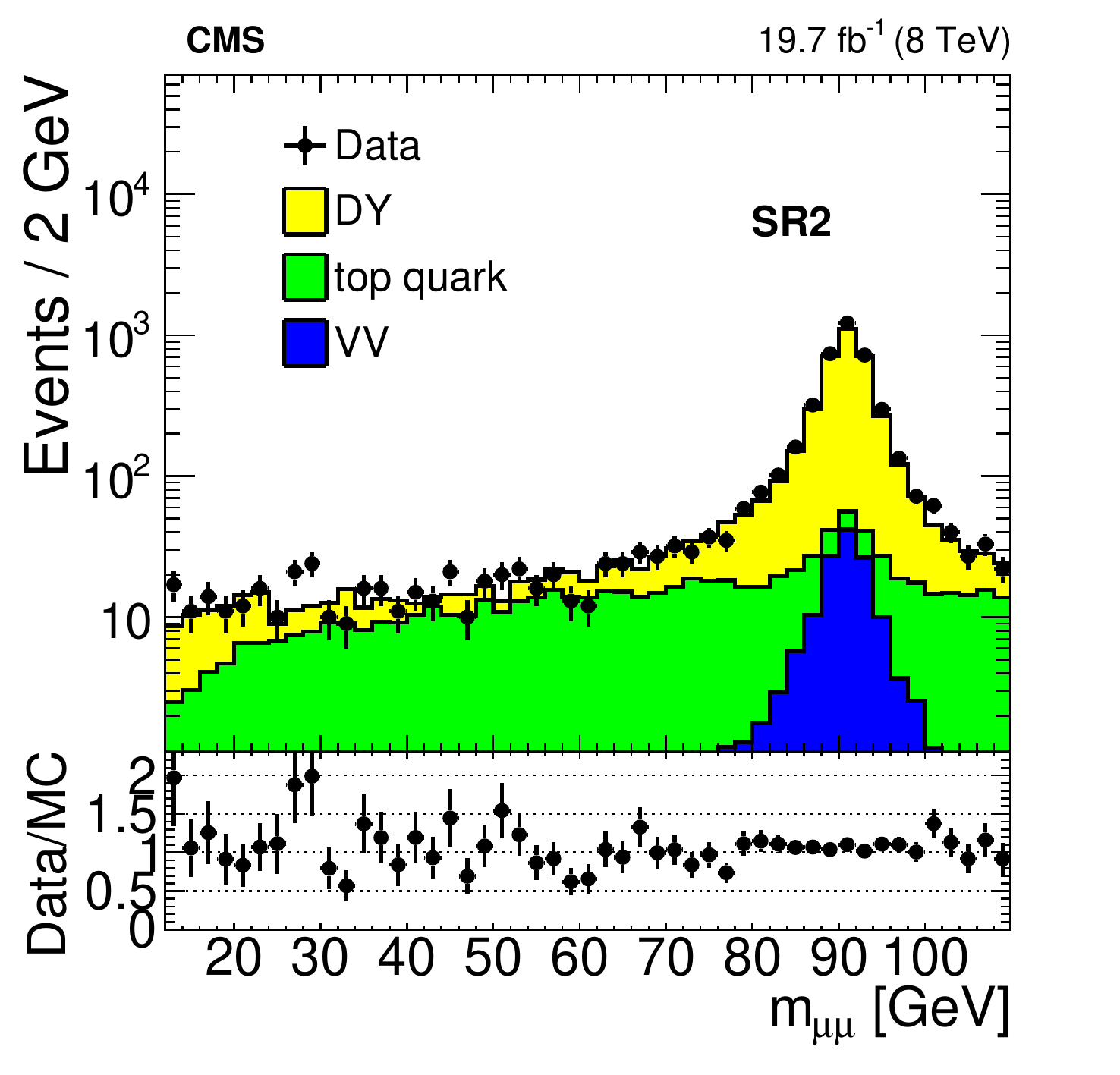}
   \includegraphics[width=0.475\textwidth]{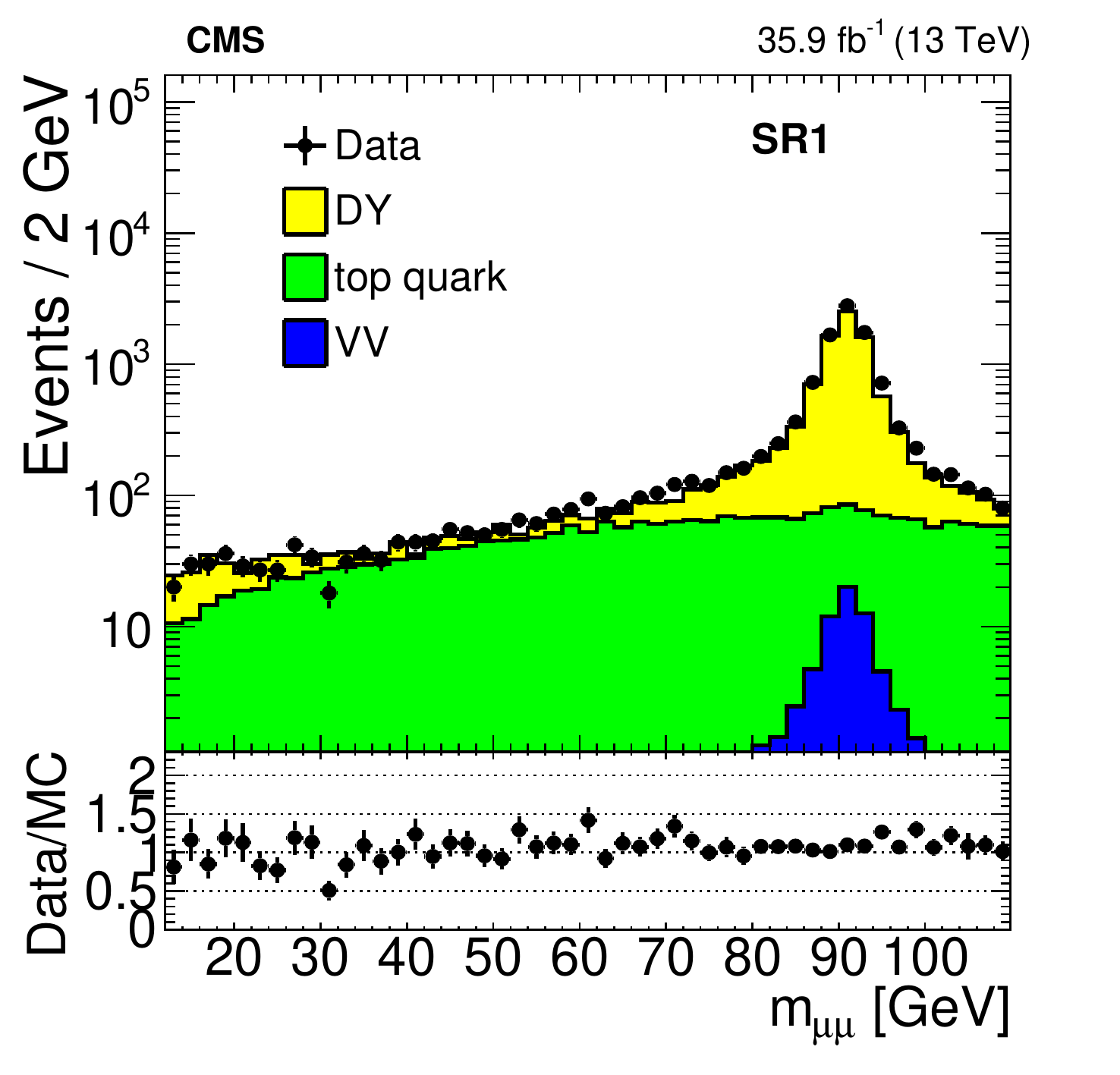}
   \includegraphics[width=0.475\textwidth]{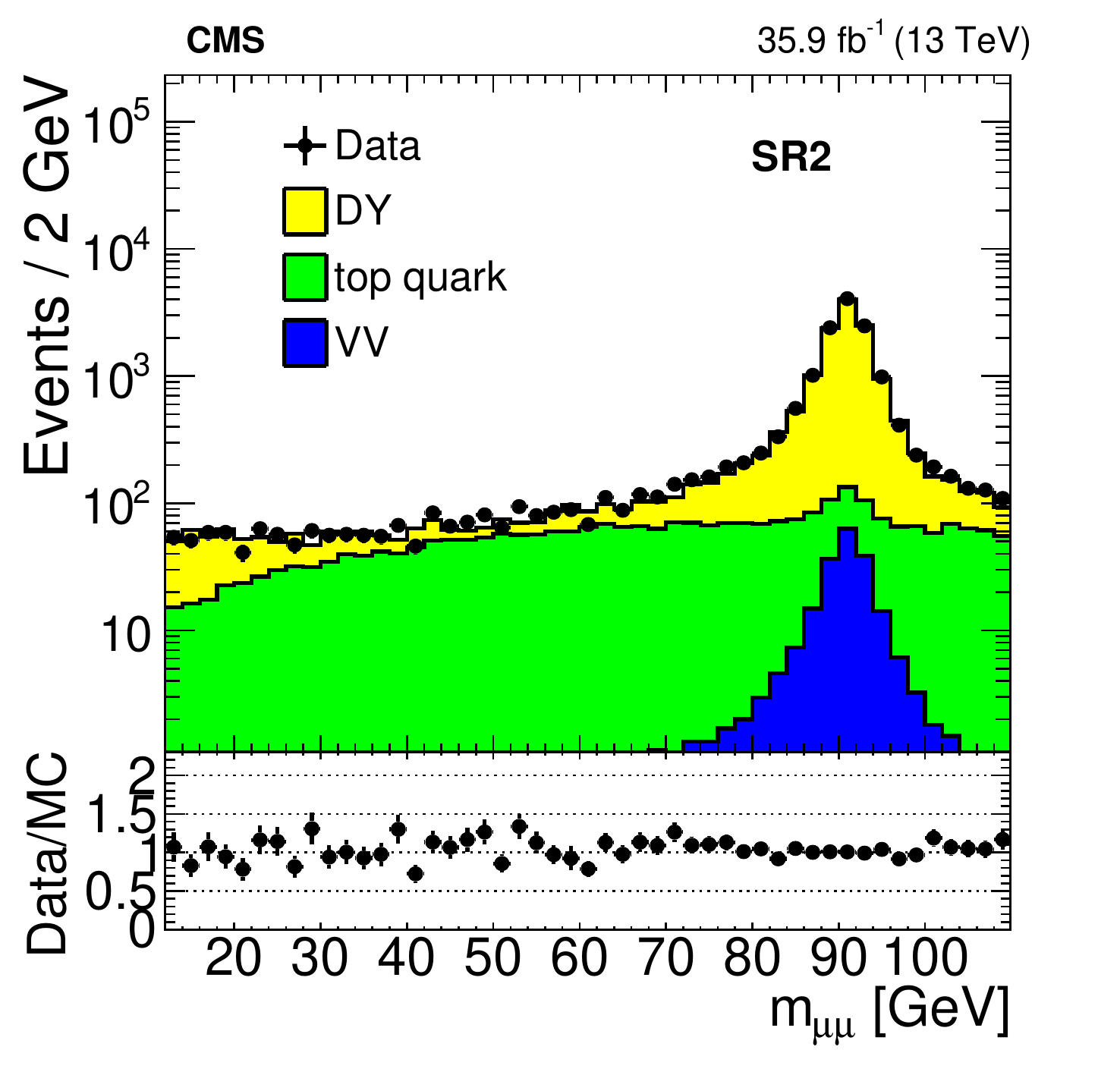}
    \caption{\label{fig:datamc}
      Upper row: the dimuon mass distribution in SR1 (left) and SR2 (right) in the 8\TeV analysis, with the
      simulation-based background expectations superimposed. Lower row: the dimuon mass distribution in SR1
      (left) and SR2 (right) in the 13\TeV analysis, with the simulation-based background expectations
      superimposed.}
    \centering
\end{figure}

The \mmm distribution for events selected in the 8\TeV data set with the SR1 requirements is shown in Fig.~\ref{fig:datamc} (upper left) compared with a simulation-based estimate of the background, dominated by the top quark events at low, and DY production at high dimuon mass. There is good agreement between data and simulation in the mass range between 12 and 24\GeV and above 34\GeV. An excess in data over the predicted background is seen in the mass range of $\simeq$26--32\GeV, which is broader than that expected from a narrow resonance.

To investigate the origin of the observed excess, we also study the dimuon mass spectrum in a complementary phase space region, SR2. It was defined from basic considerations, testing if the production process is dominated by the electroweak or the strong interaction. In the latter case the second jet may be present not in the forward but rather in the central pseudorapidity region. To compensate for an otherwise significant increase in the \ttbar background in SR2, we use additional $\ptmiss$ and $\Delta\phi(\mu\mu, \mathrm{jj})$ requirements, which are not needed in SR1. This complementary selection is also shown in Table~\ref{table:topologies}. The dimuon mass distribution in SR2 for the 8\TeV analysis is shown in Fig.~\ref{fig:datamc} (upper right) together with the background expectations from simulation. An excess is present in SR2, too, at a similar mass and with similar width.

The analysis is repeated using 13\TeV data with approximately twice the integrated luminosity of the 8\TeV sample. The \mmm distribution for events selected with the SR1 and SR2 requirements is presented in Fig.~\ref{fig:datamc} (lower left and right), together with the background expectations from simulation, and show no significant excess over the background-only hypothesis in the entire mass spectrum studied.

\section{Characterization of the dimuon mass spectra\label{sec:results}}

The \mmm spectrum is fit using a convolution of Breit--Wigner and Gaussian functions to model a possible signal where the excess is seen. The Breit--Wigner function describes the intrinsic resonance line-shape, while the Gaussian part describes the experimental mass resolution of 0.45\GeV for a dimuon system with a mass of 28\GeV. Because of a low event count in simulated background samples, a smooth polynomial function for the description of the background is used, with the parameters allowed to vary freely in the fit.

In order to characterize quantitatively any potential event excess, we perform an unbinned maximum likelihood fit to the dimuon mass distribution  \mmm in the 12--70\GeV range using the following expression for the likelihood:
\begin{linenomath}
\begin{equation}
L(\mX, \Gamma_{\mu\mu},a_{1}, a_{2}) = \frac{(\NS+\NB)^{N}}{N!}e^{-(\NS+\NB)}
\prod^N_{i=1} [\frac{\NS}{\NS+\NB} p_{i}^{\mathrm{S}}(\mX, \Gamma_{\mu\mu}) +
              \frac{\NB}{\NS+\NB} p_{i}^{\mathrm{B}}(a_{1}, a_{2})],
\end{equation}
\end{linenomath}
where $N$ is the number of observed events in data, $\NS$ is the number of the signal events, $\NB$ is the number of the background events, and $p_{i}^{\mathrm{S}}$ and $p_{i}^{\mathrm{B}}$ are the probability density functions for the signal and the background, respectively, to have a measured dimuon mass \mmm in the event $i$. The free parameters of the fit are $\NS$, $\NB$, the signal mass \mX and the width $\Gamma_{\mu\mu}$, and the parameters $a_{1}$ and $a_{2}$ of the polynomial function of the background model. The optimal choice of the order of the polynomial function for the background model (second-order for both SRs and at both center-of-mass energies) was based on the same criteria as used in the CMS SM $\PH \to \PGg \PGg$ analysis~\cite{CMS-Higgs2}.

\begin{figure}[tb!]
    \centering
\vspace{1cm}
   \includegraphics[width=0.475\textwidth]{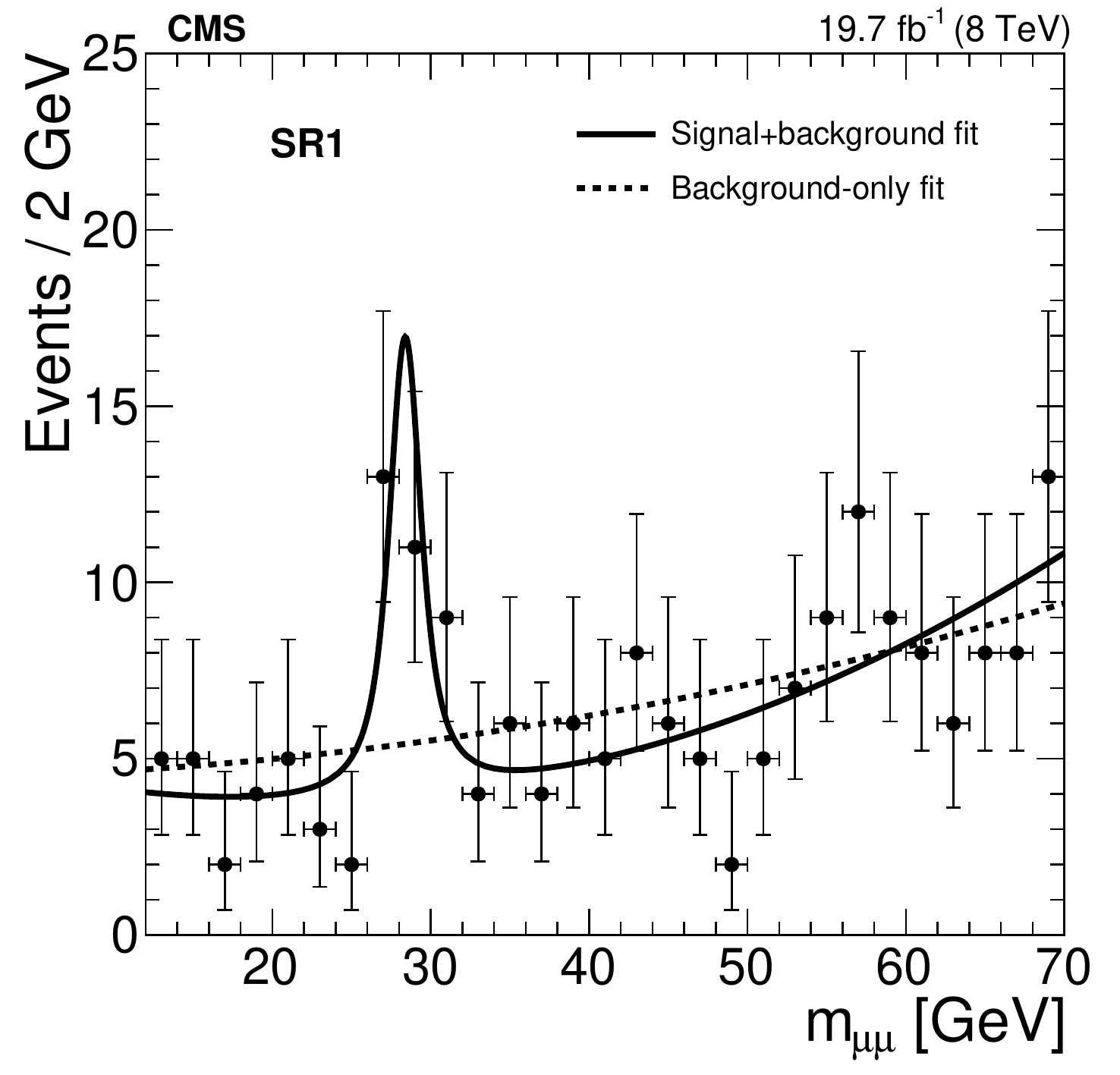}
   \includegraphics[width=0.475\textwidth]{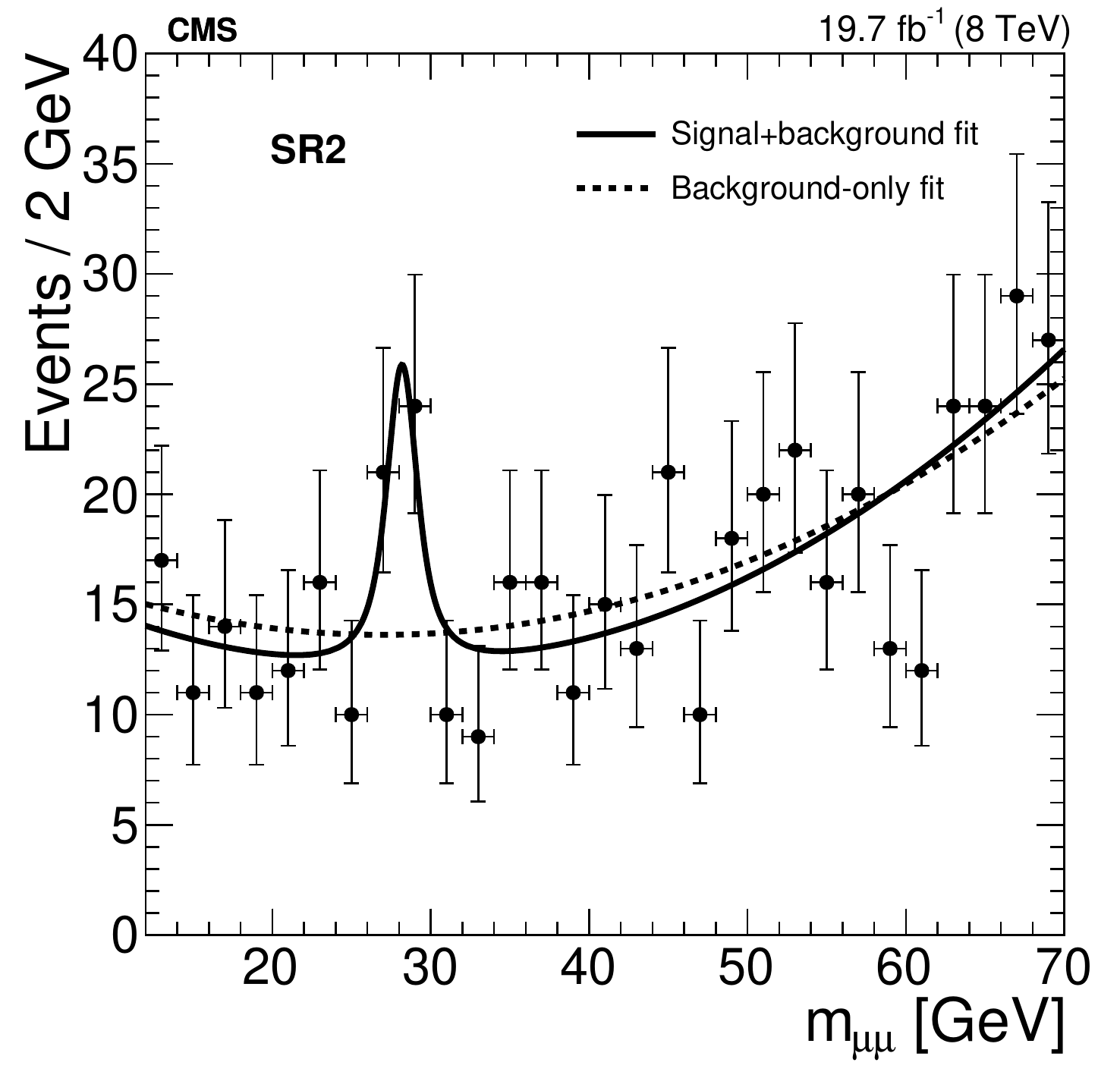}
   \includegraphics[width=0.475\textwidth]{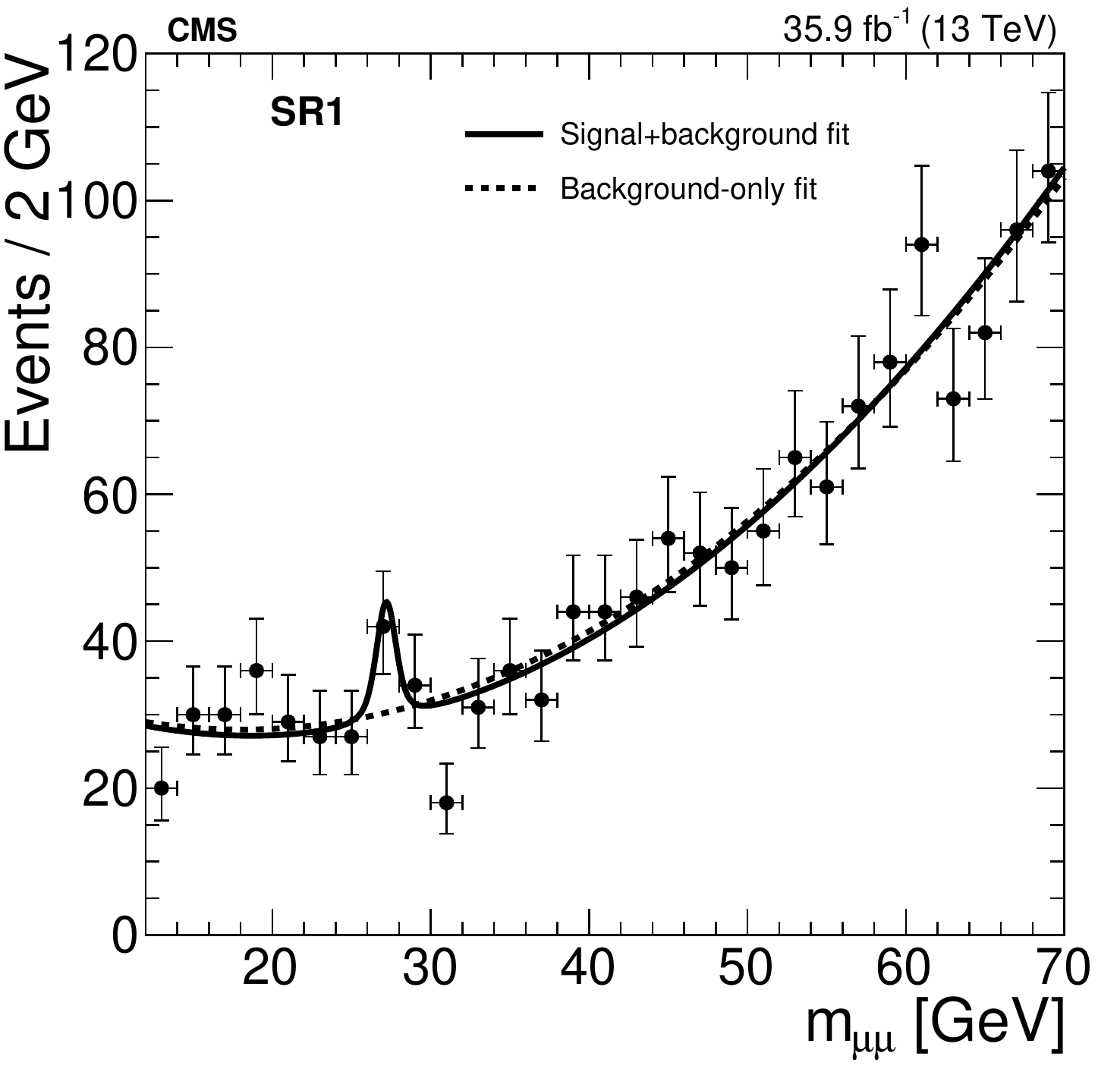}
   \includegraphics[width=0.475\textwidth]{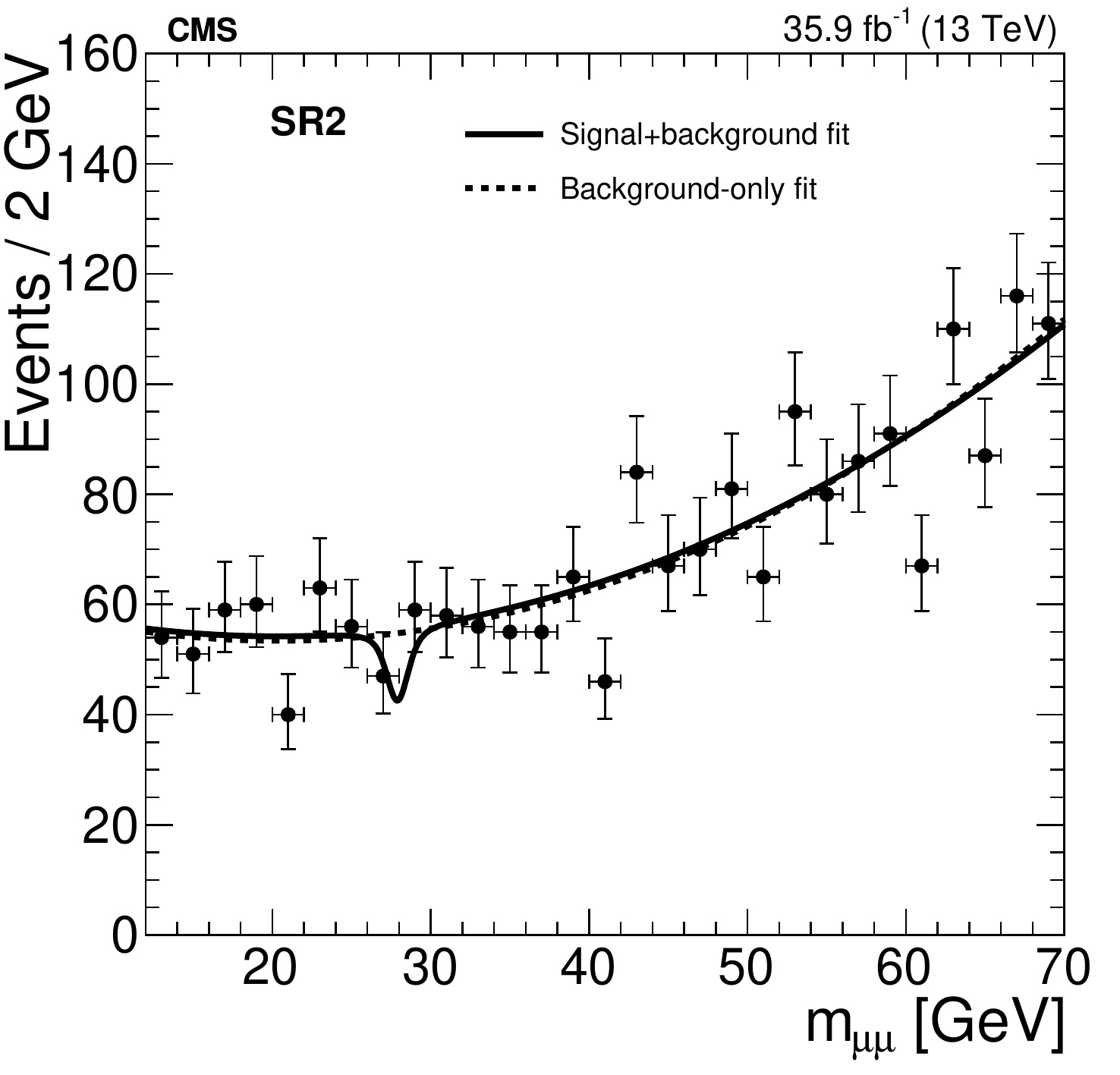}
    \caption{\label{fig:fit}
      Upper row: the $12 < \mmm < 70\GeV$ range in SR1 (left) and SR2 (right) in the 8\TeV analysis.
      Lower row: the $12 < \mmm < 70\GeV$ range in SR1 (left) and SR2 (right) in the 13\TeV analysis.
      The results of an unbinned maximum likelihood fit for the signal-plus-background (solid lines)
      and background-only (dashed lines) hypotheses are superimposed.}
    \centering
\end{figure}

The results of the fit in the $12 < \mmm < 70\GeV$ range of SR1 and SR2 for the 8\TeV analysis are shown in Fig.~\ref{fig:fit} (upper left and right). The solid line corresponds to the fit with the signal-plus-background hypothesis, while the dashed line shows the fit with the background-only hypothesis. The values of $\chi^{2}$ which characterize the agreement between the data and the fit result, are 18.5 and 22.5 for 29 bins in SR1 and SR2, respectively.

The statistical significance of the excess and the upper limits are evaluated using a
frequentist approach. A profile likelihood ratio test statistic is calculated~\cite{Cowan:2010js} as:
\begin{linenomath}
\begin{equation}
  q_A \equiv -2\,\ln \left[
  \frac  { L( \hat{m}_{\PX}, \hat{\Gamma}_{\mu\mu}, \hat{\hat{a}}_{1}, \hat{\hat{a}}_{2} ) }
         { L( \hat{m}_{\PX}, \hat{\Gamma}_{\mu\mu}, \hat{a}_1, \hat{a}_2 ) }
                  \right],
\end{equation}
\end{linenomath}
where $\hat{A}$, $\hat{\mX}$, $\hat{\Gamma}_{\mu\mu}$, $\hat{a}_1$, and  $\hat{a}_2$ are the values that maximize the likelihood $L$ given the data, and $\hat{\hat{a}}_{1}$, $\hat{\hat{a}}_{2}$ are the values that maximize the likelihood for a fixed arbitrary value of $A$. If $\hat{A} < 0$, then $q_A$ is set to zero. The evaluation of the significance of an excess is based on $q_0$, while the evaluation of an upper limit on the signal cross section is based on $q_{A}$ with $q_{A}=0$ if $\hat{A}>A$. The $q_0$ distribution for the fixed values of \mX and $\Gmm$ tends to conform to a $\chi^2$ distribution with one degree of freedom, from which the $p$-values can be calculated~\cite{Cowan:2010js}; the values obtained are verified by a large number of pseudo-experiments.

The local significance of the excess found in SR1 at 8\TeV is 4.2\unit{s.d.} A global significance of 3.0\unit{s.d.} is evaluated by taking the look-elsewhere effect (LEE)~\cite{Lyons:1900zz} into account for the given dimuon mass range and the range of the signal width 0.5--2.0\GeV. The global significance we quote does not take into account the choice of all event selection criteria, and therefore should be considered only as a partial accounting for the LEE.
The local significance of the excess observed  in SR2 at 8\TeV is 2.9\unit{s.d.} The best fit values of the
hypothetical signal mass \mX and its width $\Gamma_{\PGm\PGm}$ obtained from the fit to the 8\TeV data are
listed in Table~\ref{table:fit}.

\begin{table}[htp]
    \topcaption{The mass and width of the event excess obtained in the 8\TeV analysis.}
    \centering
    \begin{tabular}{ c  c  c } \hline
    Event & SR1 & SR2 \\
    category  & Additional forward jet  & Additional central jet \\
    \hline
\mX ({\GeVns}) &  28.4$\pm$ 0.6 & 28.2$\pm$ 0.7\\
$\Gmm$ ({\GeVns}) & 1.9$\pm$1.3  &1.9$\pm$1.1\\
\hline
    \end{tabular}
    \label{table:fit}
    \centering
\end{table}

The relative uncertainties in the muon \pt scale ($\simeq$0.2\%) and in the dimuon mass resolution ($\simeq$10\%)
have a negligible effect on the $p$-values, and the mass and width measurements.

We further perform the combined fit to the two SRs at 8\TeV to reduce the uncertainties in the extraction of the mass and the width of a hypothetical resonance. The number of the signal and background events, $\NS$ and $\NB$, and the parameters of the background functions, $a_{1}$ and $a_{2}$, in the two SRs are varied independently in the fit, while the common signal mean and width are used in both SRs. The mean and the width of the signal extracted from the combined fit are $\mathrm{m}_{\PX}=28.3\pm0.4\GeV$ and $\Gmm=1.8\pm0.8\GeV$.

Several cross-checks are performed to evaluate the stability of the observed excess in the 8\TeV analysis. The analysis is repeated using an alternative jet reconstruction algorithm~\cite{CMS-PAS-JME-14-005}; using a double-muon, instead of the single-muon, trigger; with alternative kinematic selections  targeting a reduction of the dominant \ttbar background (increased $\ptmiss$ requirement, the use of the variable
$\mT=\sqrt{\smash[b]{(\pt^{\mu1})^2+(\pt^{\mu2})^2+(\ptmiss)^2}}$
instead of the $\ptmiss$ selection, and a change in the jet veto threshold). In all cases we observe a statistically significant excess with the local significance within 0.5\unit{s.d.} of that for the nominal selections. We also checked that the event excess is observed with relaxed or tighter \PQb tagging selections and after dropping either the muon isolation or pileup jet identification criteria.

A similar analysis of the dimuon mass spectrum in 13\TeV data shows no significant excess near 28\GeV in either SR1 or SR2. Figure~\ref{fig:fit} (lower left) shows the dimuon mass spectrum in the $12 < \mmm < 70\GeV$ range for events in SR1, with the fit result superimposed. The fit yields a mild excess corresponding to a local significance of 2.0\unit{s.d.}, with a fitted mass of $27.2 \pm 0.6\GeV$ and a width of $0.7 \pm 1.0\GeV$. Figure~\ref{fig:fit} (lower right) shows the dimuon mass spectrum in SR2 together with the fit result, which yields a negative signal yield with a significance of 1.4\unit{s.d.} The corresponding $\chi^2$ values are 21.0 and 36.5 for SR1 and SR2, respectively.

We provide a measurement of the fiducial cross sections and upper limits at 95\% confidence level (\CL) on those for a potential signal. The limits are obtained under the background-only hypothesis and using an asymptotic approximation~\cite{Cowan:2010js} of the \CLs method~\cite{Junk:1999kv,Read:2002hq}. The quoted values take into account the reconstruction efficiency $\varepsilon^{\mathrm{reco}}$, which includes the muon trigger, identification, and isolation efficiency, as well as the {\cPqb} tagging efficiency. It was obtained from simulation using the \ttbar sample with the dimuon decays of the top quark pairs, which is the dominant background in the mass region of the search. In the absence of a reliable model predicting a hypothetical signal, it is not possible to include the efficiency of the kinematic selections. Consequently, the fiducial cross section is reported, defined as:
\begin{linenomath}
\begin{equation}
  \sigma_{\mathrm{fid}} = \frac{\NS}{\lumi \,\varepsilon^{\mathrm{reco}}},
\end{equation}
\end{linenomath}
where $\NS$ is the number of the signal events extracted from the fit, \lumi is the integrated luminosity, and $\varepsilon^{\mathrm{reco}}$ is the reconstruction efficiency. The relative uncertainties in the muon trigger, identification, and isolation efficiency (3\%), the {\cPqb} tagging efficiency (1.6\% at 8\TeV and 1.0\% at 13\TeV), and the integrated luminosity measurement (2.6\% at 8\TeV~\cite{CMS-PAS-LUM-13-001} and 2.5\% at 13\TeV~\cite{CMS-PAS-LUM-17-001}) are taken into account in the fit as nuisance parameters. For 8\TeV data a combined fit in the two SRs is performed and these uncertainties are considered as fully correlated between SR1 and SR2. The effect of the systematic uncertainties is negligible compared to the statistical uncertainty. The values of the signal mass and the width, and their associated uncertainties obtained from the combined fit to the 8\TeV data in the two SRs, are used in the fit of the 13\TeV data, which is performed separately for each SR.

Table~\ref{table:sigmas} shows the local significances, the mass and the width of the event excess, the measured fiducial cross sections with $\pm1$\unit{s.d.} uncertainties, and the 95\% \CL upper limits on those. The best fit $\NS$ values for the two SRs and the 95\% \CL upper limits on those, the reconstruction efficiencies and the integrated luminosities are also listed.

\begin{table}[htp]
    \topcaption{The local significances, the mass and the width of the event excess, the measured fiducial cross sections with $\pm1$\unit{s.d.} uncertainties, and the 95\% \CL upper limits on those. The best fit $\NS$ values for the two SRs and the 95\% \CL upper limits on those, the reconstruction efficiencies and the integrated luminosities are also listed.}
    \centering
    \begin{tabular}{ccccc} \hline
    $\sqrt{s}$ ({\TeVns})        &       \multicolumn{2}{c}{8}       &   \multicolumn{2}{c}{13} \\
    Event category                        & SR1       &  SR2      & SR1              & SR2  \\
    \hline
    Local significance (s.d.)    &          4.2         & 2.9       & 2.0              & 1.4 deficit \\
    \mX ({\GeVns})               & \multicolumn{2}{c}{28.3$\pm$ 0.4} & 27.2$\pm$0.6   & \\
    $\Gmm$ ({\GeVns})            & \multicolumn{2}{c}{1.8$\pm$0.8}   & 0.7$\pm$1.0    & \\
    $\NS$                        &    $22.0 \pm7.6$ & $22.8 \pm 9.5$ & $14.5 \pm9.3$ & $-14.9\pm 10.1$ \\
    $\NS$ observed upper limit at 95\% \CL  & $40.4$          & $44.7$      & $36.9$     & $32.2$ \\
    $\NS$ expected upper limit at 95\% \CL       & $18.3$          & $27.6$      & $27.6$     & $35.6$ \\
    $\varepsilon^\mathrm{reco}$   & \multicolumn{2}{c}{$0.27 \pm 0.01$} & \multicolumn{2}{c}{$0.28 \pm 0.01$} \\
   Integrated luminosity, \lumi (fb$^{-1}$) & \multicolumn{2}{c}{$19.7 \pm 0.5$} & \multicolumn{2}{c}{$35.9 \pm 0.9$} \\
    $\sigma_\mathrm{fid}$ (fb)             & $4.1 \pm 1.4$  & $4.2 \pm 1.7$  & $1.4 \pm 0.9$ & $-1.5 \pm 1.0$ \\
    Observed upper limit at 95\% \CL (fb)           & $7.6$  & $8.4$  & $3.7$ & $3.2$  \\
    Expected upper limit at 95\% \CL (fb)           & $3.4$  & $5.2$  & $2.7$ & $3.5$  \\
    \hline
    \end{tabular}
    \label{table:sigmas}
    \centering
\end{table}

\begin{figure}[htp]
    \centering
\vspace{1cm}
   \includegraphics[width=0.48\textwidth]{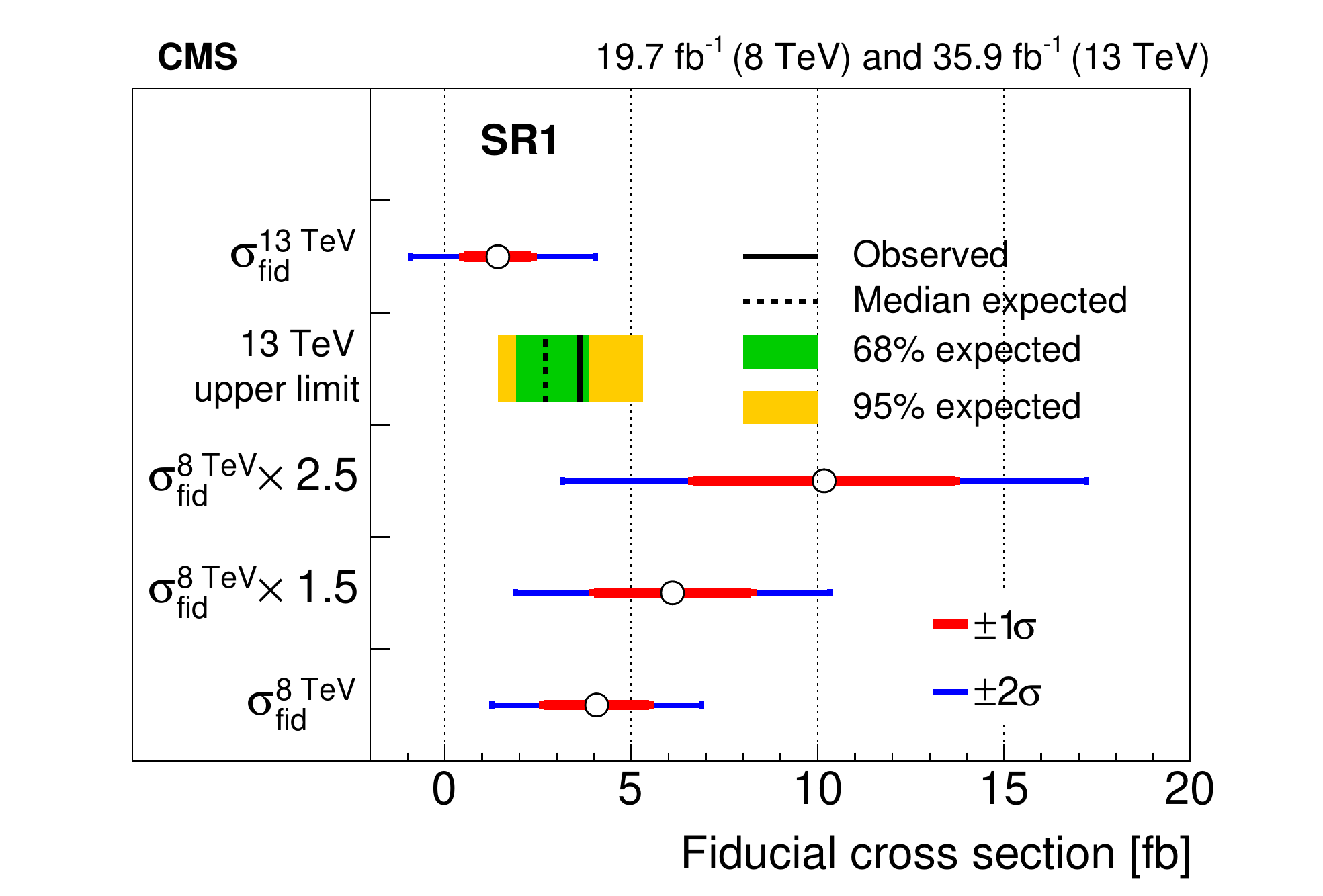}
   \includegraphics[width=0.48\textwidth]{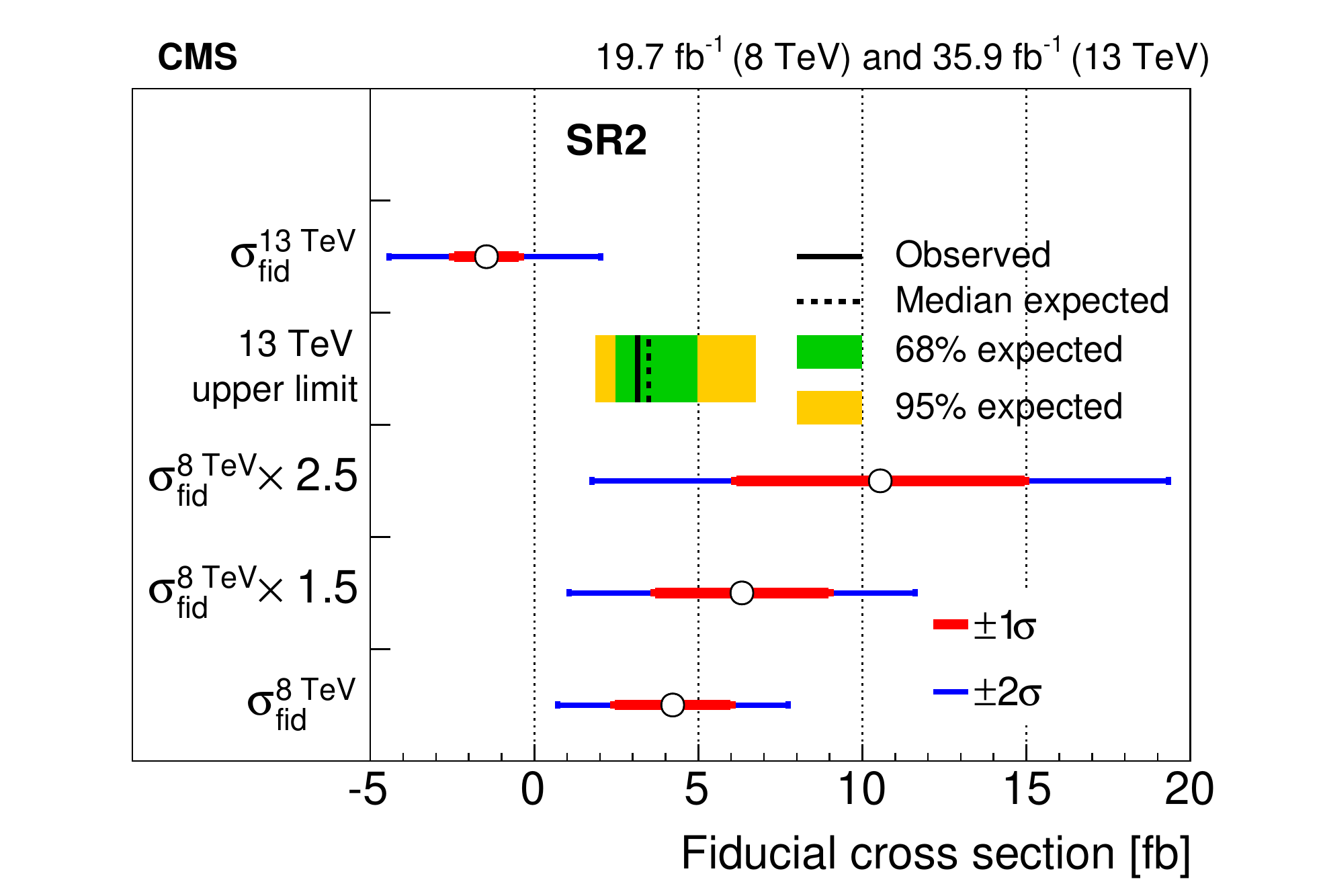}
    \caption{\label{fig:summary}
     The measured fiducial signal cross sections and the 95\% \CL upper limits on those
     in SR1 (left) and SR2 (right). The expected (observed) upper limits are shown
     as vertical dashed (solid) lines, together with the 68 and 95\% \CL uncertainties in the expected limits
     (under the background-only hypothesis). Also shown are the expected 13\TeV cross sections and their
     uncertainties obtained by scaling the measured 8\TeV cross sections by the factors of 1.5 and 2.5,
     as discussed in the text.}
    \centering
\end{figure}

Figure~\ref{fig:summary} presents a summary of the measured fiducial cross sections and the 95\% \CL upper limits on those
in SR1 (left) and SR2 (right). The expected (observed) upper limits are shown as vertical dashed (solid) lines, together with the 68 and 95\% \CL uncertainties in the expected limits. Also shown in the plot are the expected cross sections and their uncertainties at $\sqrt{s}=13\TeV$, which were obtained by scaling the measured 8\TeV cross sections by a factor of 1.5 or 2.5, indicative of an expected cross section increase from $\sqrt{s}= 8$ to 13\TeV for the $\PQq\PQq$ or $\Pg\Pg$ production mechanism, respectively, for the invariant mass of the produced system in the mass range between 30\GeV and the top quark mass~\cite{Bertone:2013vaa,Carrazza:2014gfa}. The choice of the lower edge of this range is motivated by the measured mass of a hypothetical dimuon resonance. The upper edge was taken assuming that a hypothetical resonance could be produced in a top quark decay. In the absence of a realistic signal model, both the mass range and the scaling should be considered only as simple benchmarks; in particular, the scaling does not take into account possible changes in the signal acceptance between the two collision energies; hence we can not exclude that the signal kinematics seen with the 8\TeV selections are disfavored in 13\TeV data.

We note that the event excess at 8\TeV cannot be explained by a light pseudoscalar Higgs boson produced in association with a \PQb quark pair, $\Pp\Pp \to \bbbar \PSA$, $\PSA \to \PGmp\PGmm$. Even assuming $\sigma(\bbbar\PSA)\mathcal{B}(\PSA \to \PGt^+\PGt^-)$ as large as 100\unit{pb} for $m_\PSA = 30\GeV$, attainable in the wrong-sign Yukawa coupling scenario in the 2HDM~\cite{Bernon:2014nxa}, the expected number of signal events after the selection is too small if the $\PSA\to\PGmm\PGmp$ branching fraction is obtained as $\mathcal{B}(\PSA\to\PGmm\PGmp)=(m_\PGm/m_\PGt)^2 \mathcal{B}(\PSA\to\PGt^+\PGt^-)$. Neither can the event excess be explained by the processes $\Pg\Pg \to \PH(125)\to  \PSA\PSA \to  \PGmp\PGmm\bbbar$ or $\Pg\Pg \to \Ph_2 \to \Ph_1 \Ph_1 \to  \PGmp\PGmm\bbbar$ in 2HDM or NMSSM (in the case where $\Ph_2$ is identified with the $\PH(125)$ boson), which also yield too low cross sections when taking into account various existing theoretical and experimental constraints. We note that the above statement also holds when the (potentially negative) interference effects between these two processes are taken into account, as well as all possible additional contributions from $\qqbar \to \PGmp\PGmm \PAQb\PQq$ + c.c. electroweak and QCD diagrams (where $\PQq$ can be either a \PQb or a light quark), none of which yields a significant cross section enhancement.

\section{Summary\label{sec:summary}}

We report on a search for resonances in the mass range 12--70\GeV, produced in association with a \PQb quark jet and another jet, and decaying to a muon pair. The analysis is based on data from proton-proton collisions at center-of-mass
energies of 8 and 13\TeV, collected with the CMS detector at the LHC and corresponding to integrated luminosities of 19.7 and 35.9\fbinv, respectively. The search is carried out in two mutually exclusive event categories. Events in the first category are required to
have a \PQb quark jet in the central region ($\abs{\eta}\le 2.4$) and at least one jet in the forward region
($\abs{\eta}>2.4)$. Events in the second category are required to have two jets in the
central region, at least one of which is identified as a \PQb quark jet, no jets in the forward region, and low missing transverse momentum. An excess of events above the background near a dimuon mass of 28\GeV is observed in both event categories in the 8\TeV data, corresponding to local significances of 4.2 and 2.9 standard deviations, respectively.

A mild excess of data over the background in the first event category is observed in 13\TeV data and corresponds to a local significance of 2.0 standard deviations, while the second category results in a deficit with a local significance of 1.4 standard deviations.

We provide a measurement of the fiducial cross sections and the upper limits on those at 95\% confidence level, evaluated for the mass and the width values obtained from the combined fit to the two event categories in $\sqrt{s}=8\TeV$ data. In the lack of a realistic signal model, the 13\TeV results are not sufficient to make a definitive statement about the origin of the 8\TeV excess. Therefore, more data and additional theoretical input are both required to fully understand the results presented in this paper.

\begin{acknowledgments}
We congratulate our colleagues in the CERN accelerator departments for the excellent performance of the LHC and thank the technical and administrative staffs at CERN and at other CMS institutes for their contributions to the success of the CMS effort. In addition, we gratefully acknowledge the computing centers and personnel of the Worldwide LHC Computing Grid for delivering so effectively the computing infrastructure essential to our analyses. Finally, we acknowledge the enduring support for the construction and operation of the LHC and the CMS detector provided by the following funding agencies: BMWFW and FWF (Austria); FNRS and FWO (Belgium); CNPq, CAPES, FAPERJ, FAPERGS, and FAPESP (Brazil); MES (Bulgaria); CERN; CAS, MoST, and NSFC (China); COLCIENCIAS (Colombia); MSES and CSF (Croatia); RPF (Cyprus); SENESCYT (Ecuador); MoER, ERC IUT, and ERDF (Estonia); Academy of Finland, MEC, and HIP (Finland); CEA and CNRS/IN2P3 (France); BMBF, DFG, and HGF (Germany); GSRT (Greece); NKFIA (Hungary); DAE and DST (India); IPM (Iran); SFI (Ireland); INFN (Italy); MSIP and NRF (Republic of Korea); MES (Latvia); LAS (Lithuania); MOE and UM (Malaysia); BUAP, CINVESTAV, CONACYT, LNS, SEP, and UASLP-FAI (Mexico); MOS (Montenegro); MBIE (New Zealand); PAEC (Pakistan); MSHE and NSC (Poland); FCT (Portugal); JINR (Dubna); MON, RosAtom, RAS, RFBR, and NRC KI (Russia); MESTD (Serbia); SEIDI, CPAN, PCTI, and FEDER (Spain); MOSTR (Sri Lanka); Swiss Funding Agencies (Switzerland); MST (Taipei); ThEPCenter, IPST, STAR, and NSTDA (Thailand); TUBITAK and TAEK (Turkey); NASU and SFFR (Ukraine); STFC (United Kingdom); DOE and NSF (USA).

\hyphenation{Rachada-pisek} Individuals have received support from the Marie-Curie program and the European Research Council and Horizon 2020 Grant, contract No. 675440 (European Union); the Leventis Foundation; the A. P. Sloan Foundation; the Alexander von Humboldt Foundation; the Belgian Federal Science Policy Office; the Fonds pour la Formation \`a la Recherche dans l'Industrie et dans l'Agriculture (FRIA-Belgium); the Agentschap voor Innovatie door Wetenschap en Technologie (IWT-Belgium); the F.R.S.-FNRS and FWO (Belgium) under the ``Excellence of Science - EOS" - be.h project n. 30820817; the Ministry of Education, Youth and Sports (MEYS) of the Czech Republic; the Lend\"ulet (``Momentum") Program and the J\'anos Bolyai Research Scholarship of the Hungarian Academy of Sciences, the New National Excellence Program \'UNKP, the NKFIA research grants 123842, 123959, 124845, 124850 and 125105 (Hungary); the Council of Science and Industrial Research, India; the HOMING PLUS program of the Foundation for Polish Science, cofinanced from European Union, Regional Development Fund, the Mobility Plus program of the Ministry of Science and Higher Education, the National Science Center (Poland), contracts Harmonia 2014/14/M/ST2/00428, Opus 2014/13/B/ST2/02543, 2014/15/B/ST2/03998, and 2015/19/B/ST2/02861, Sonata-bis 2012/07/E/ST2/01406; the National Priorities Research Program by Qatar National Research Fund; the Programa Estatal de Fomento de la Investigaci{\'o}n Cient{\'i}fica y T{\'e}cnica de Excelencia Mar\'{\i}a de Maeztu, grant MDM-2015-0509 and the Programa Severo Ochoa del Principado de Asturias; the Thalis and Aristeia programs cofinanced by EU-ESF and the Greek NSRF; the Rachadapisek Sompot Fund for Postdoctoral Fellowship, Chulalongkorn University and the Chulalongkorn Academic into Its 2nd Century Project Advancement Project (Thailand); the Welch Foundation, contract C-1845; and the Weston Havens Foundation (USA).
\end{acknowledgments}
\bibliography{auto_generated}
\cleardoublepage \appendix\section{The CMS Collaboration \label{app:collab}}\begin{sloppypar}\hyphenpenalty=5000\widowpenalty=500\clubpenalty=5000\input{HIG-16-017-authorlist.tex}\end{sloppypar}
\end{document}

%% file: HIG-16-017-authorlist.tex
\vskip\cmsinstskip
\textbf{Yerevan Physics Institute, Yerevan, Armenia}\\*[0pt]
A.M.~Sirunyan, A.~Tumasyan
\vskip\cmsinstskip
\textbf{Institut f\"{u}r Hochenergiephysik, Wien, Austria}\\*[0pt]
W.~Adam, F.~Ambrogi, E.~Asilar, T.~Bergauer, J.~Brandstetter, M.~Dragicevic, J.~Er\"{o}, A.~Escalante~Del~Valle, M.~Flechl, R.~Fr\"{u}hwirth\cmsAuthorMark{1}, V.M.~Ghete, J.~Hrubec, M.~Jeitler\cmsAuthorMark{1}, N.~Krammer, I.~Kr\"{a}tschmer, D.~Liko, T.~Madlener, I.~Mikulec, N.~Rad, H.~Rohringer, J.~Schieck\cmsAuthorMark{1}, R.~Sch\"{o}fbeck, M.~Spanring, D.~Spitzbart, A.~Taurok, W.~Waltenberger, J.~Wittmann, C.-E.~Wulz\cmsAuthorMark{1}, M.~Zarucki
% ...
\vskip\cmsinstskip
% ...
1:  Also at Vienna University of Technology, Vienna, Austria\\
% ...
